 \documentclass[onecollarge,natbib]{svjour2}

 \usepackage{graphicx}
\usepackage{epsfig}

\def\nl{\noindent}

\def\sla#1{\rlap\slash #1}

\newcommand{\be}{\begin{equation}}
\newcommand{\ee}{\end{equation}}
\newcommand{\cc}{\c c}

\newcommand{\bee}{\begin{eqnarray}}
\newcommand{\eee}{\end{eqnarray}}

\bibpunct{[}{]}{;}{n}{}{,} 
\smartqed  
\usepackage{graphicx,amsmath,amssymb}
\usepackage{color}
\usepackage{enumerate,array}

\usepackage{multicol}

\definecolor{navyblue}{rgb}{0.3,0.3,1}
\definecolor{purple}{rgb}{0.6,0,0.5}
\usepackage[colorlinks=true, pdfstartview=FitV, 
linkcolor=purple, citecolor= navyblue, urlcolor=navyblue]{hyperref}

\journalname{Few Body Systems}
\begin{document}

\title{
Eletroweak Form Factors  
 in the Light-Front for Spin-1 Particles
\footnote{Presented by J.~P.~B.~C.~de Melo at LIGHTCONE 2011,23 -27 May, 2011 Dallas.}}  
\author{J.~P.~B.~C.~de~Melo \and T.~Frederico}

\institute{J.~P.~B.~C.~de~Melo  \at
                Laboratorio de F\'isica 
Te\'orica e Computa\cc\~ao Cient\'ifica, 
Universidade Cruzeiro do Sul, 01506-000, S\~ao Paulo, Brazil. \\
                \and T.~Frederico \at
                Departamento de F\'isica, 
Instituto Tecnol\'ogico de Aeron\'autica, CTA, 
12228-900, Sa\~o Jos\'e dos Campos, Brazil.}

\date{Version of \today}

\maketitle

\begin{abstract}
The contribution of the light-front valence wave function to the
electromagnetic current of spin-1 composite particles is not
enough to warranty the proper transformation of the current under
rotations. The naive derivation of the plus component of the
current in the Drell-Yan-West frame within an analytical and
covariant model of the vertex leads to the violation of the
rotational symmetry. Computing the form-factors in a quasi
Drell-Yan-West frame $q^+\rightarrow 0$, we were able to separate
out in an analytical form the contributions from Z-diagrams or
zero modes using the instant-form cartesian polarization basis. 
\end{abstract}

\section{Introduction}
\label{intro}
Light-front models are useful to describe hadronic bound states,
like mesons or baryons due to its particular boost
properties~\cite{Terentev76,Brodsky98}. However, the light-front
description in a truncated Fock-space breaks the rotational
symmetry because the associated transformation is a dynamical
boost~\cite{Pacheco97,Pacheco98,Pacheco992,sales2000}. 
Therefore, an analysis with
covariant analytical models, can be useful to pin down  the main
missing features in a truncated light-front Fock-space description
of the composite system. In this respect, the rotational symmetry
breaking of the plus component of the electromagnetic current, in
the Drell-Yan frame, was recently studied within an analytical
model for the spin-1 vertex of a composite two-fermion bound
state~\cite{Pacheco97,Pacheco98}. 

It was shown that, if pair term
contributions are ignored in the evaluation of the matrix elements
of the electromagnetic current, the covariance of the form factors
is lost~\cite{Pacheco97,Pacheco98,Pacheco992,JI2001,Choi2004}.

The complete
restoration of covariance in the form factor calculation is found
only when pair terms or zero modes contributions to the matrix
elements of the current are
considered~\cite{Pacheco98,Pacheco992,Choi2004,Naus98,Pacheco99}.

The extraction of the electromagnetic form-factors of a spin-1
composite particle from the microscopic matrix elements of the
plus component of the current ($J^+=J^0+J^3$) in the Drell-Yan
frame (momentum transfer $q^+=q^0+q^3=0$), based only on the
valence component of the wave function, presents ambiguities due
to the lacking of the rotational invariance of the current
model~\cite{Inna84,Inna89}. In the Breit-frame with  momentum
transfers along the transverse direction (the Drell-Yan condition
is satisfied) the current $J^+$ has four independent matrix
elements, although only three form factors exist. Therefore, the
matrix elements satisfies an identity, known as the angular
condition~\cite{Inna84}, which is violated.

Several extraction schemes for evaluating the form-factors were
proposed, and in particular we consider the suggestion made in
Ref.~\cite{Inna84}. It was found in a numerical calculation  of the
$\rho$-meson electromagnetic form factors considering only the
valence contribution~\cite{Pacheco97}, that the prescription
proposed by~\cite{Inna84} to evaluate the form-factors, produced
results in agreement with the covariant calculations. In
Ref.~\cite{Pacheco97}, it was used an analytical form of the
$\rho$-quark-antiquark vertex.  Later, in Ref.~\cite{JI2001}, it was
shown that the above prescription eliminates the pair diagram
contributions to the form factors, using a simplified form of the
model, when the matrix elements of the current were evaluated for
spin-1 light-cone polarization states. This nice result was thought
to be due to the use of the particular light-cone polarization
states. Here, we will show in a straightforward and analytic manner
that the cancelation of the pair contribution in the evaluation of
the form factors using the prescription from Ref.~\cite{Inna84} also
happens for the instant form polarization states in the cartesian
representation, generalizing the previous conclusion~\cite{JI2001}.
Our aim, is to expose in a simple and detailed form, how the pair
terms appear in the matrix elements of the current evaluated between
instant form polarization states, and their cancelation in the form
factors using the correct prescription. Therefore, we conclude that
this property is more general than realized before.
But, in the case of spin-0 composite particles
(like the pion) with the correspondent form of the analytical
model, the plus component of the electromagnetic current in the
Breit-frame, with $q^+=0$, does not have contributions from pair
terms~\cite{Pacheco99,Pacheco2002}. 

\section{Light-Front Model Spin-1 Particles and Wave Function}

The electromagnetic current has the following general form for
 spin-1 particles~\cite{Frankfurt79}:
 \begin{equation}
 J_{\alpha \beta}^{\mu}=[F_1(q^2)g_{\alpha \beta} -F_2(q^2)
 \frac{q_{\alpha}q_{\beta}}{2 m_v^2}] (p^\mu + p^{\prime \mu})
  - F_3(q^2)
 (q_\alpha g_\beta^\mu- q_\beta g_\alpha^\mu) \ ,  
\label{eq:curr1}
 \end{equation}
 where $m_v$ is the mass of the vector particle, $q^\mu$ is the
 momentum transfer, $p^\mu$ and $p^{\prime  \mu}$ is on-shell initial
 and final momentum respectively.  From the covariant form factors
 $F_1$, $F_2$ and $F_3$, one can obtain the charge ($G_0$), magnetic
 ($G_1$) and quadrupole ($G_2$) form factors (see e.g.~\cite{Pacheco97}).

The matrix elements of the electromagnetic current ${\cal J}
_{ji}={\epsilon^\prime_j}^\alpha \epsilon^\beta_iJ_{\alpha
\beta}^{\mu}$ in the impulse approximation are written
as~\cite{Pacheco97}:
\begin{equation}
{\cal J}^+_{ji} =  \imath  \int\frac{d^4k}{(2\pi)^4}
 \frac{ Tr[\epsilon^{'\nu}_j \Gamma_{\beta}(k,k-p_f)
(\sla{k}-\sla{p_f} +m)  
\gamma^{+} 
(\sla{k}-\sla{p_i}+m) \epsilon^\mu_i \Gamma_{\alpha}(k,k-p_i)
(\sla{k}+m)]
\Lambda(k,p_f)\Lambda(k,p_i) }
{(k^2 - m^2+\imath \epsilon)  
((k-p_i)^2 - m^2+\imath\epsilon) 
((k-p_f)^2 - m^2+\imath \epsilon )} 
\end{equation}
where ${\epsilon^\prime_j}$ and ${\epsilon_i}$ are the
polarization four-vectors of the final and initial states,
respectively and $m$ is the quark mass. 
The electromagnetic form-factors are calculated in the Breit frame
with the Drell-Yan-West condition, which gives the momentum transfer
$q^\mu=(0,q_x,0,0)$, the particle initial momentum
$p^\mu=(p^0,-q_x/2,0,0)$ and the the final one
$p^{\prime\mu}=(p^0,q_x/2,0,0)$. We use $\eta=-q^2/4 m_{v}$ and
$p^0=m_{v}\sqrt{1+\eta}$. The polarization four-vectors in
instant-form basis are given by 
$\epsilon^{\mu}_x =(-\sqrt{\eta},\sqrt{1+\eta},0,0),~~
\epsilon^{\mu}_y=(0,0,1,0),~~  \epsilon^{\mu}_z=(0,0,0,1),$ 
for the initial state and by, 
$\epsilon^{\prime \mu}_x=(\sqrt{\eta},\sqrt{1+\eta},0,0),~~
\epsilon^{\prime\mu}_y=\epsilon^{\mu}_y ,~~
\epsilon^{\prime\mu}_z=\epsilon^{\mu}_z,$ 
for the final state. 
The regularization function
$\Lambda(k,p_{i(f)}) = N/((p-k)^2-m_R +\imath \epsilon)^2$ 
is enough to render finite the photo-absorption amplitude; 
the regularization parameter is $m_R$. 
The vertex function $m_v-q\bar{q}$ 
for the vector particles 
utilized~(see ref.~\cite{Pacheco97} for details) is 
given below:
\begin{equation}
\Gamma^\mu (k,p) = \gamma^\mu -\frac{m_v}{2}
\frac{2 k^\mu -p^\mu}
{ p.k + m_{v} m -\imath \epsilon}   \ .
\label{rhov}
\end{equation}
The vector particle is on-mass shell; 
$m_v$ is the vector bound state mass, and  
its four momentum $p^{\mu}=k^{\mu} - k^{\prime \mu}$.  
After the integration in the light-front energy 
($k^-=(\vec{k}^2_{\perp} + m^2)/k^+$),  
the light-front wave function is writing like:
\begin{equation}
\Phi_i(x,\vec k_\perp)=
\frac{N^2}{(1-x)^2(m^2_v - M_0^2)
(m^2_v - M^2_R)^2} 
\vec \epsilon_i . [\vec \gamma -  \frac{\vec k}
{\frac{M_0}{2}+ m}]  \ . 
\label{eq:npwf}
\end{equation} 
The extraction of the electromagnetic form factor 
with the plus component of the electromagnetic current 
and the angular condition are discussed next section.

\section{Light-Front Spin Basis and the Angular Condition}

The matrix elements of the electromagnetic current 
expressed in terms of the current in the instant form spin basis, 
after the use of the Melosh 
rotation~\cite{Pacheco97,Keister91,Frankfurt93} is: 
\begin{eqnarray}
I^{+}_{11} & = & \frac{J^{+}_{xx}+(1+\eta) J^{+}_{yy}-
\eta J^{+}_{zz}  - 2 \sqrt{ \eta} J^{+}_{zx}}{2 (1+\eta)}
\  , \  
I^{+}_{10} = \frac{\sqrt{2 \eta} J^{+}_{xx} + 
\sqrt{2 \eta} J^{+}_{zz}
 - \sqrt{2} (\eta-1) J^{+}_{zx}}{2(1+\eta)}
\nonumber \\
I^{+}_{1-1}& = &\frac{-J^{+}_{xx}+(1+\eta) J^{+}_{yy}+
\eta J^{+}_{zz} +  2 \sqrt{\eta} J^{+}_{zx}}{2 (1+\eta)}
\ , \ 
I^{+}_{00} = \frac{-\eta J^{+}_{xx}+J^{+}_{zz} 
 -  2 \sqrt{\eta} J^{+}_{zx}}
{(1+\eta)} \ , 
\label{eq:ifront1}
\end{eqnarray}
here the matrix elements of the electromagnetic current are 
calculated with the plus component of the current, ''$+$'' and the 
light-front polarizatiion states denoted as $I^+_{m^{\prime}m}$. 

In the Breit-frame ($q^+=0$), 
the angular condition  is translated by the equation
~\cite{Pacheco97,Keister91}:
\begin{equation}
\Delta(q^2)=(1+2 \eta) I^{+}_{11}+I^{+}_{1-1} - 
\sqrt{8 \eta} I^{+}_{10} -
I^{+}_{00} \ = \ (1 + \eta)(J^+_{yy}-J^+_{zz})=0  \ .
\label{eq:ang}
\end{equation} 
In the case of the instant form spin basis, the angular condition is 
$J^{+}_{yy}=J^{+}_{zz}$~\cite{Frankfurt93}.  
The prescription adopted by Grach and Kondratyuk~\cite{Inna84} 
eliminate the matrix elements $I^+_{00}$ in order to compute the 
electromagnetic form factors for spin-1 particles.

The terms of the trace calculated with only the 
$\gamma^{\mu}$ structure of the vertex, 
of the Eq.(\ref{rhov}), is writing below:
\begin{equation}
 Tr[gg]_{ji}= Tr\left[ 
\sla{\epsilon}_j^{\alpha} (\sla{k}-\sla{p}^{\prime} + m) 
\gamma^+ (\sla{k} - \sla{p} + m) \sla{\epsilon}_i^{\alpha}
(\sla{k} + m)
\right] \ .
\label{trgaga}
\end{equation} 
The trace is calculated with the light-front coordinates,
(~$k^+=k^0+k^3,~k^-=k^0-k^3,~k_{\perp}=(k_x,k_y)$) 
and the $k^-$ 
dependence is separate,( in the Eq.(\ref{trgaga})), then, we 
get the results below:
\begin{equation}
 Tr[gg]_{ji}^{+Z} = \frac{k^-}{2} 
Tr\left[ 
\sla{\epsilon}_j^{\alpha} (\sla{k}-\sla{p}^{\prime} + m) 
\gamma^+ (\sla{k} - \sla{p} + m) \sla{\epsilon}_i^{\alpha}
\gamma^+  \right]  \ .
\end{equation}  
For the polarization four-vectors
$(\epsilon_i^{\mu},\epsilon_j^{\mu})$, the traces above are given by:
\begin{eqnarray}
 Tr[gg]^{+Z}_{xx}
& = & - k^- \frac{\eta}{8} R_{gg} \ , \ 
Tr[ gg ]_{yy}^{+Z}  =  \  4 k^- (p^+ -k^+)^2 \ ,
\nonumber \\
Tr[gg]_{zz}^{+Z} & = & \frac18 ~k^- ~R_{gg} \ , \ 
Tr[gg]^{+Z}_{zx} =  -  k^- \frac{ \sqrt{\eta} }{8} R_{gg}
\label{traces} \ ,
\end{eqnarray}
where $R_{gg}= 4~Tr[ (\sla{k}-\sla{p^\prime} +m) \gamma^{+}
(\sla{k}-\sla{p}+m) \gamma^-] 
\label{badtrace}
\ . $
The trace equation,~Eq.(\ref{traces}), yield the following
relations:
\begin{eqnarray}
 Tr[ gg ]^{+Z}_{xx}
& = & -\eta Tr[ gg ]_{zz}^{+Z} ~,   \nonumber \\
Tr[ gg ]^{+Z}_{zx} & = &  -\sqrt{\eta}~Tr[ gg ]^{+Z}_{zz}  \ .
\label{trrelations}
\end{eqnarray}

The zero-modes contributions for the direct term coupling 
given by:
\begin{equation}
J_{ji}^{+Z}[gg] =
 \lim_{\delta^+ \rightarrow 0_+}\;\sum_{r=4,5\; ;\;s=4,6}
\int [d^4 k]^{+Z} \frac{ Tr[gg ]^{+Z}_{ji} } {
[1] [2] [3 ] [r ]^2 [ s ]^2} \ , 
\label{currimgg}
\end{equation}
where $[d^4 k]^Z=[d^4 k]~\theta(p^{\prime+}-k^+) \theta(k^+-p^{+})$,
and the denominators are given by: $ [1] =
[k^2-m^2+\imath\epsilon]$,
 $[2 ]  =   [(k-p)^2-m^2 + \imath \epsilon]$,
 $[ 3 ] =   [(k-p^{\prime })^2 -  m^2 +\imath
\epsilon]$, $[ 4 ]  =  [(k-p)^2-m_R^2 + \imath \epsilon]$, 
$[ 5 ] = [(k-p^{\prime })^2 -  m_R^2 +\imath \epsilon]$, 
$[ 6 ]  = [k^2- m_R^2 +\imath \epsilon]$ and $p^{\prime+}=p^+ + \delta^+$.
The integration of the Eq.(\ref{currimgg}) is performed with the 
{\it pole dislocation method}, developed in the reference~\cite{Naus98}, 
after that, 
the contribution of the {\it zero modes} for the matrix element 
$J^{+Z}_{yy}$ of the 
electromagnetic current is zero and the matrix element 
$J^{+}_{yy}$ of the electromagnetic current not have a  
pair term contribuition.

Using the prescription of the Ref.\cite{Inna84}, 
the electromagnetic form factors for spin-1 particles 
are writing with matrix elements of the electromagnetic current in 
instant form spin basis as:
\begin{eqnarray} 
G_0^{GK} & = &  \frac{1}{3} \left( J_{xx}^{+} + \eta
J_{zz}^{+}+ (2- \eta)J_{yy}^{+}\right) \ , \ G_1^{GK} = J^+_{yy}
-\frac{1}{\sqrt{\eta}}\left( J_{zx}^{+}+\sqrt{\eta}J_{zz}^{+}
\right) , \nonumber  \\
G_2^{GK}  & = &   \frac{\sqrt{2}}{3} \left( J_{xx}^{+}+ \eta J_{zz}^{+} -
(1+ \eta)J_{yy}^{+}  \right) \ ,  
\label{ffactors}
\end{eqnarray}
here the {\it superscripts GK} in the equation above 
is the initials for the 
authors of the Ref.~\cite{Inna84},~Grach,~I.~ and Kondratyuk, L.~A.

Considering the relations in Eq.(\ref{trrelations}), the prescription 
given by Grach~\cite{Inna84}, cancel out the zero-modes contribution 
to the electromagnetic form factors for this part of the vertex function
~(see the figures 1 and 2). 
Then the angular condition in the Cartesian spin basis with the zero-modes 
contribuition is~\cite{Pacheco97}:
\begin{equation}
\Delta(q^2)=J^+_{yy}-J^+_{zz}=
J^{+val}_{yy} - J^{+val}_{zz} -J^{+Z}_{zz}=0  \ ,
\label{acondition}
\end{equation}
because the zero modes contribuition to the 
matrix elements $J^{+Z}_{yy}$ of the electromagnetic 
current is zero; 
the angular condition is only due to the valence components 
of the electromagnetic current, $J^{+val}_{yy}, J^{+val}_{zz}$, 
and to the zero modes contribuition to the 
electromagnetic current~$J^{+Z}_{zz}$.

The figures 1 and 2 show the results for the electromagnetic 
matrix elements of the current calculated with the 
instant form and the light-front approach. After the zero-modes 
inclusion, the covariance are restorate.

\begin{figure}[tbh!]
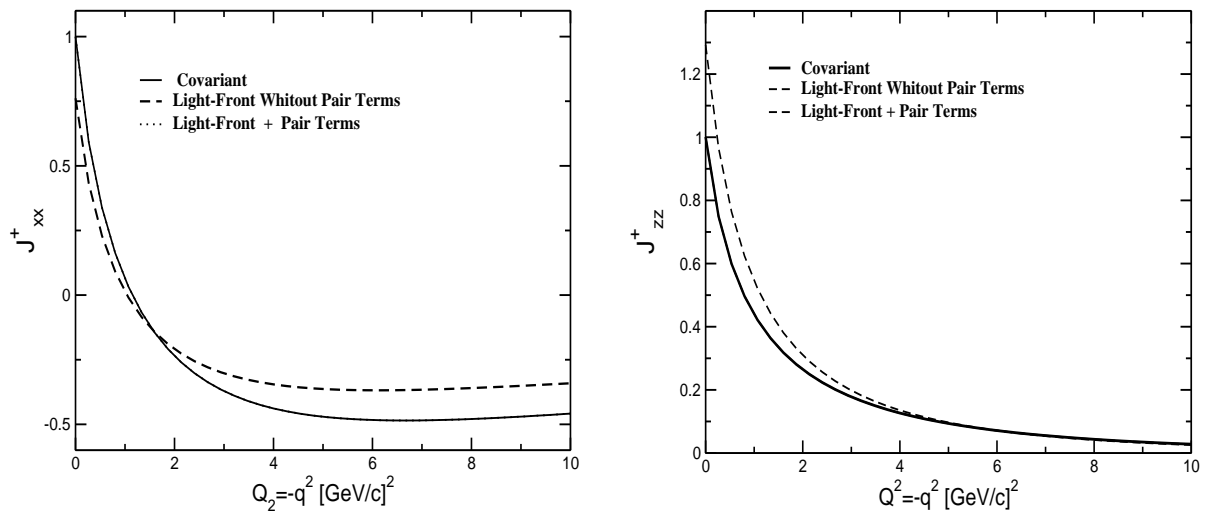

\centerline{
\epsfig{figure=cuxxgg11.eps,width=7.50cm,height=7.50cm} 
\hspace{0.4444000cm} 
\epsfig{figure=cuzzgg11.eps,width=7.50cm,height=7.50cm} 
}
\caption{The plots show the matrix elements of the electromagnetic current, 
$J^+_{xx}$ and $J^+_{zz}$, 
calculated with the coupling $(\gamma^{\mu},\gamma^{\nu})$ 
in the light-front approach and the instant form calculation; 
the parameters utilized are $m=0.430~GeV$, $m_{v}=0.770~GeV$
and the regulator mass as $m_{R}=~1.8~GeV$.}
\label{fig12}
\end{figure}

\begin{figure}[tbh!]
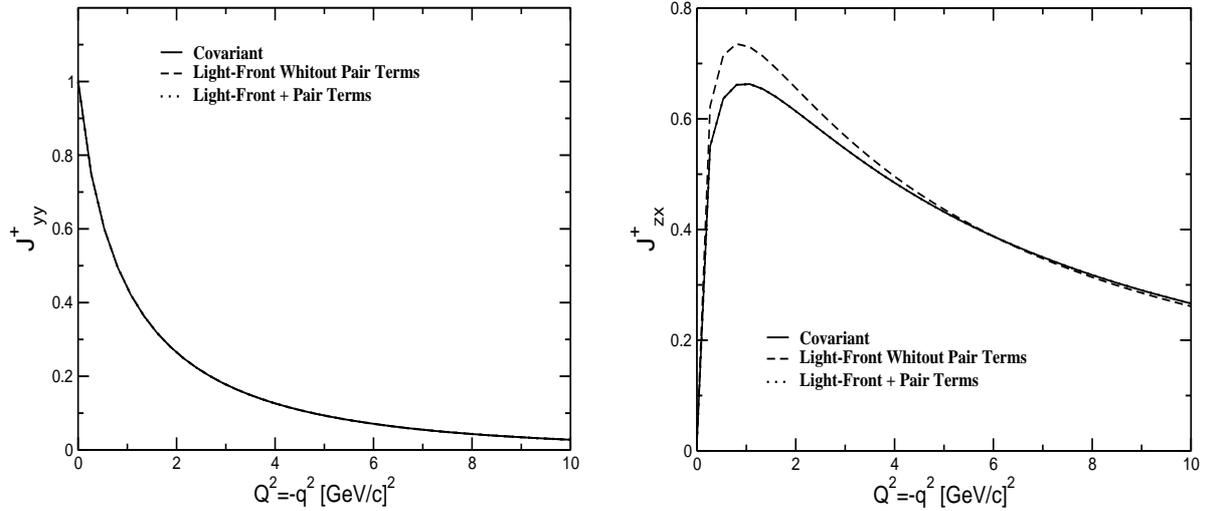

\centerline{
\epsfig{figure=cuyygg11.eps,width=7.50cm,height=7.50cm}
 \hspace{0.4444000cm} 
\epsfig{figure=cuzxgg11.eps,width=7.50cm,height=7.50cm}  
}
\caption{The plots show the matrix elements of the electromagnetic current, 
$J^+_{yy}$ and $J^+_{zx}$, 
calculated with the coupling $(\gamma^{\mu},\gamma^{\nu})$ 
in the light-front approach and the instant form calculation;
  the parameters utilized are $m=0.430~GeV$, $m_{v}=0.770~GeV$
and the regulator mass as $m_{R}=~1.8~GeV$.}
\label{fig34}
\end{figure}

\section{Conclusions}

Following the Ref.~\cite{Pacheco97} we separate 
the structure of the vertex~Eq.(\ref{rhov}) in 
the valence contribuition and the zero-modes for the 
spin-1 particles. 

The prescription sugested by the 
Ref.~\cite{Inna84} utilized in the calculation of 
the electromagnetic form factor 
for spin-1 particles cancel out the zero-modes contribuition to the 
matrix elements for the electromagnetic current; then the 
electromagnetic form factors for spin-1 calculated with the 
{\it Grach} and {\it Kondratyuk}~\cite{Inna84} 
are free of the non-valence contribuition and only the valence 
part of the electromagnetic current contributed 
to the electromagnetic form factors. 

It is possible to generalize the conclusion found here 
for the full electromagnetic 
matrix elements of the spin-1 particles with the vertex 
function, Eq.(\ref{rhov}); i.e, the zero-modes not contributed to the 
electromagnetic form factors if the prescription of the 
Ref.\cite{Inna84} it is adopted.

\vspace{0.2cm}

\nl {\bf Acknowledgments.} This work was supported in part by the
Brazilian agencies FAPESP (Funda\c{c}\~ao de Amparo \`a Pesquisa do
Estado de S\~ao Paulo) and CNPq (Conselho Nacional de
Desenvolvimento Cient\'\i fico e Tecnol\'ogico). 
J. P. B. C. de Melo thanks the organizer 
 of the Light-Cone 2011 for the 
 invitation. 


 \end{document}